\documentclass{svproc}
\usepackage{url}
\usepackage{hyperref}
\usepackage{booktabs}
\usepackage{float}
\usepackage{graphicx}
\usepackage[numbers]{natbib}
\usepackage{adjustbox}
\usepackage{rotating}
\usepackage{placeins}
\usepackage{subfig}
\usepackage{cite}
\usepackage[table,xcdraw]{xcolor}

\begin{document}
\mainmatter              
\title{Can Deep Learning Reliably Recognize Abnormality Patterns on Chest X-rays? \\ A Multi-Reader Study Examining One Month of AI Implementation in Everyday Radiology Clinical Practice}
\titlerunning{Can DL Reliably Recognize Abnormality Patterns on CXR?}  
%
\author{Daniel Kvak\inst{1,2} \and Anna Chromcová\inst{1} \and Petra Ovesná\inst{3} \and Jakub Dandár\inst{4} Marek Biroš\inst{1,5} \and Robert Hrubý\inst{1,6} \and Daniel Dufek\inst{1} \and Marija Pajdaković\inst{1,7}}
\authorrunning{Kvak, Chromcová, Ovesná, Dandár, Biroš, Hrubý, Dufek \& Pajdaković} 
%
\tocauthor{Daniel Kvak et al.}
\institute{Carebot s.r.o., Prague, Czech Republic\\
\email{daniel.kvak@carebot.com}\\ 
\and
Faculty of Medicine, Masaryk University, Brno, Czech Republic
\and
Institute of Biostatistics and Analysis, s.r.o., Brno, Czech Republic
\and
Havířov Hospital and Clinic, Havířov, Czech Republic
\and
Faculty of Mathematics and Physics, Charles University, Prague, Czech Republic
\and
Faculty of Nuclear Sciences and Physical Engineering, Czech Technical University, Prague, Czech Republic
\and
Faculty of Electrical Engineering, Czech Technical University, Prague, Czech Republic}

\maketitle              

\begin{abstract}
In this study, we utilize a deep-learning-based automatic detection algorithm (DLAD, Carebot AI CXR) to detect and localize seven specific radiological findings (atelectasis (ATE), consolidation (CON), pleural effusion (EFF), pulmonary lesion (LES), subcutaneous emphysema (SCE), cardiomegaly (CMG), pneumothorax (PNO)) on chest X-rays (CXR). We collected 956 CXRs and compared the performance of the DLAD with that of six individual radiologists who evaluated the images in a simulated hospital setting. The proposed DLAD achieved high sensitivity (ATE 1.000 (0.624-1.000), CON 0.864 (0.671-0.956), EFF 0.953 (0.887-0.983), LES 0.905 (0.715-0.978), SCE 1.000 (0.366-1.000), CMG 0.837 (0.711-0.917), PNO 0.875 (0.538-0.986)), even when compared to the radiologists (\textit{LOWEST}: ATE 0.000 (0.000-0.376), CON 0.182 (0.070-0.382), EFF 0.400 (0.302-0.506), LES 0.238 (0.103-0.448), SCE 0.000 (0.000-0.634), CMG 0.347 (0.228-0.486), PNO 0.375 (0.134-0.691), \textit{HIGHEST}: ATE 1.000 (0.624-1.000), CON 0.864 (0.671-0.956), EFF 0.953 (0.887-0.983), LES 0.667 (0.456-0.830), SCE 1.000 (0.366-1.000), CMG 0.980 (0.896-0.999), PNO 0.875 (0.538-0.986)). The findings of the study demonstrate that the suggested DLAD holds potential for integration into everyday clinical practice as a decision support system, effectively mitigating the false negative rate associated with junior and intermediate radiologists.

\keywords{Artificial Intelligence; Computer-Aided Detection; Deep Learning; Chest X-ray; Radiology.}
\end{abstract}

\section{Introduction}
Despite advances in imaging technologies such as computed tomography (CT) and magnetic resonance imaging (MRI), X-ray remains a key diagnostic method thanks to its affordability, rapid scanning time, and widespread use across hospitals. Chest X-ray (CXR) is used for routine health check-ups, pre-operative examinations, screening programs, and diagnosis of cardiopulmonary diseases. However, analysis of chest radiographs can be challenging as it requires careful investigation of complex structures, there is a risk of overlooking abnormalities, and changes may appear similar in different pathologies or one pathology may show different features. This leads to the potential for errors, with Donald \& Barnard 2012 \citep{donald2012common} reporting that up to 22\% of all diagnostic radiological errors are made when interpreting chest radiographs. 

\section{Background}
Computer-aided detection (CAD) applications offer a potential solution to address the challenges in medical diagnosis by assisting physicians in improving accuracy and efficiency. The advancement of machine learning (ML) techniques in biomedical imaging has facilitated the transition of CAD tools from research settings to clinical applications. Approved systems utilizing artificial intelligence (AI) and deep learning (DL) methods have undergone validation and are expected to undergo further enhancements. Notably, DL-based applications have demonstrated promising outcomes in medical image analysis, such as the detection of diabetic retinopathy in eye images \citep{qummar2019deep}, segmentation of breast cancer \citep{al2018fully}, and identification of metastases in pathological samples \citep{janowczyk2016deep}.

In our previous studies, we have explored deep learning algorithms for the detection of suspicious lung parenchymal lesions \citep{kvak2023chest} and COVID-19 disease \citep{kvak2022towards} on CXRs. However, the clinical utility of these algorithms was limited as CXRs may exhibit various abnormalities beyond malignant nodules or pneumonia. To ensure the usefulness of a CAD system, it should be capable of processing CXRs presenting a range of abnormalities, particularly the prevalent chest diseases found within the population.

\subsection{Literature Review Methodology}
A comprehensive literature search was conducted with the objective of gathering pertinent clinical data pertaining to the medical device under investigation. The primary goal was to provide an overview of the current state of the art, including a detailed description of the clinical context and identification of potential risks associated with the device. The search encompassed both positive and negative findings to ensure a comprehensive evaluation of the proposed DLAD.
\newpage

\begin{table}[H]
\centering
\begin{tabular}{|l|l|l|l|}
\hline
\textbf{Database} & \textbf{Keywords}                                                                                                           & \textbf{\begin{tabular}[c]{@{}l@{}}Number of \\ search results\end{tabular}} & \textbf{\begin{tabular}[c]{@{}l@{}}Number of \\ studies used\end{tabular}} \\ \hline
PubMed            & \begin{tabular}[c]{@{}l@{}}artificial intelligence, deep learning, \\ detection, chest x-ray, chest radiograph\end{tabular} & 537                                                                          & 15                                                                         \\ \hline
\end{tabular}
\caption{\label{tab:search}Search results.}
\end{table}

For analysis of the related works, we utilized the PubMed database as our primary source, employing specific keywords such as "artificial intelligence", "deep learning", "detection", "chest x-ray", and "chest radiograph" (\autoref{tab:search}). Considering the limitations and lack of reliability associated with publicly available datasets \citep{hryniewska2021checklist} for constructing robust clinical models \citep{oakden2020exploring}, we excluded studies that relied on these datasets for training or testing purposes. From a total of 537 relevant studies, 15 studies involving devices that have successfully undergone certification according to Regulation (EU) 2017/745 of the European Parliament and of the Council of 5 April 2017 on medical devices were selected for analysis, following the guidelines of MEDDEV 2.7/1 rev. 4.

\subsection{Related Works}
A study conducted by Ahn et al. 2022 \citep{ahn2022association} assessed the diagnostic accuracy of detecting four specific findings: pneumonia, pulmonary nodule, pleural effusion, and pneumothorax. Another commercially available computer-aided detection (CAD) system was utilized by Park et al. 2020 \citep{park2020deep} to detect multiple classes of lesions, including nodules/mass, interstitial opacity, pleural effusion, and pneumothorax. Singh et al. 2018 \citep{singh2018deep} focused on the detection and analysis of various abnormalities such as cardiac shadow enlargement, pleural effusion, pulmonary opacities, and hilum prominence. The detection of pulmonary diseases was also addressed in the study by Sung et al. 2021 \citep{sung2021added}. Jones et al. 2021 \citep{jones2021assessment} evaluated the performance of a deep learning-based CAD system in detecting acute findings. Hwang et al. 2019 (a) \citep{hwang2019development} presented a deep learning-based CAD system for classifying normal versus abnormal findings associated with lung cancer, active pulmonary tuberculosis, pneumonia, or pneumothorax. Kim et al. 2021 \citep{kim2021effect} focused on the sensitivity of the CAD system in detecting abnormal findings. Another study by Hwang et al. 2019 (b) \citep{hwang2019development} specifically targeted the detection of active tuberculosis, followed by studies conducted by Lee et al. 2021 \citep{lee2021deep} and Nash et al. 2020 \citep{nash2020deep}. Qin et al. 2019 \citep{qin2019using} evaluated tuberculosis detection in the context of identifying TB-related abnormalities and comparing the performance of three different deep learning-based CAD systems. Jang et al. 2020 \citep{jang2020deep} focused solely on the detection of pneumonia using a deep learning-based CAD system. The reproducibility of test-retest for the detection of pulmonary nodules was analyzed by Kim et al. 2020 \citep{kim2020test}, and this topic was also addressed in the studies by Koo et al. 2021 \citep{koo2021extravalidation} and Nam et al. 2019 \citep{nam2019development}. In studies employing a multi-reader design, specifically comparing general practitioners (GPs) with radiologists, it was found that the use of the algorithm improved the efficiency of reporting findings, thereby enhancing physician performance, as concluded by multiple studies.

\section{Software}
\begin{figure}[H]
\centering
\includegraphics[width=1\textwidth]{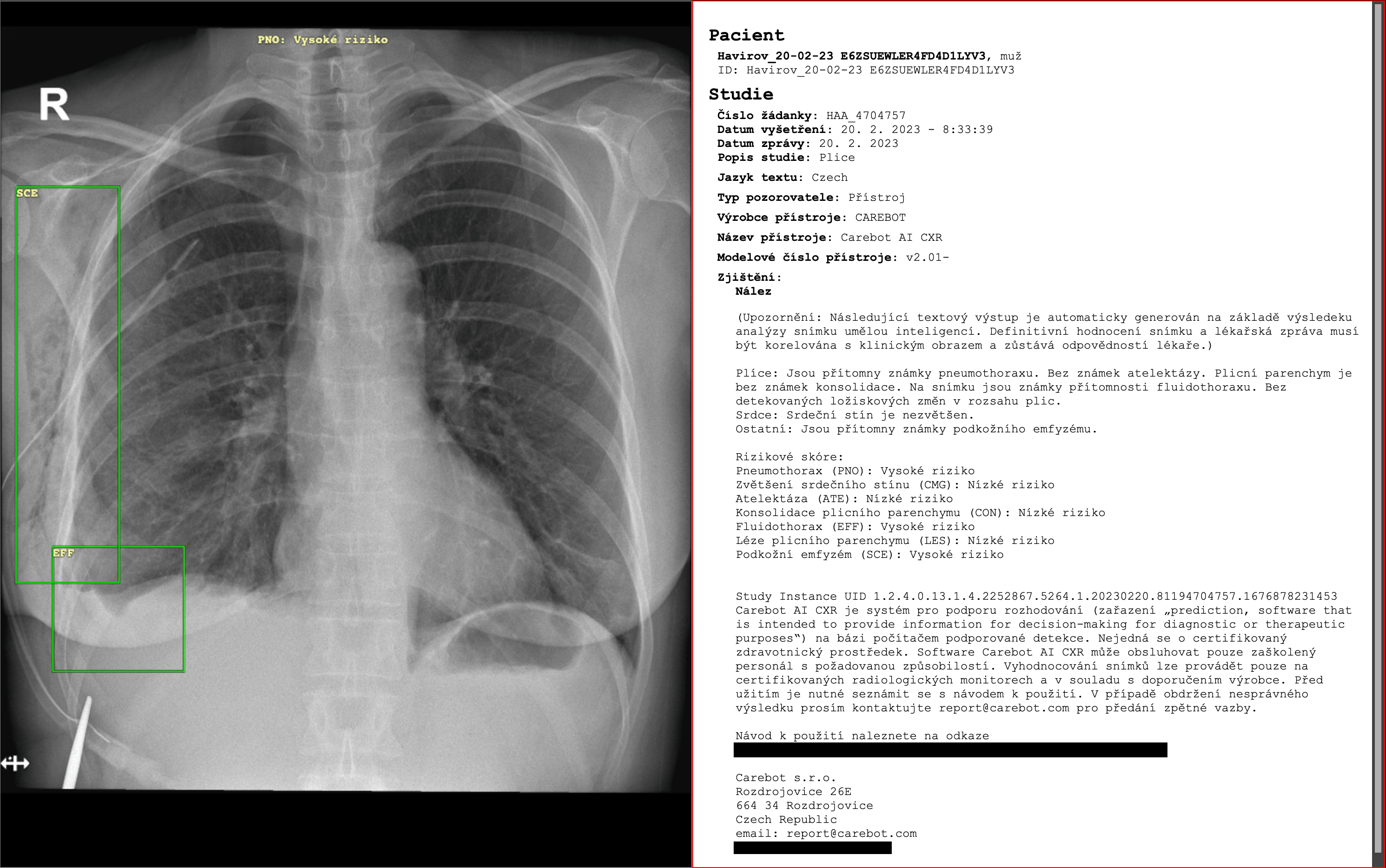}
\caption{\label{fig:showcase}\textbf\scriptsize{A demonstration of the proposed DLAD (Carebot AI CXR) implemented in the picture archiving and communication system (OR-CZ CloudPACS). The DICOM Presentation State (shown as green boxes) generated by the DLAD as an overlay of the patient's original study indicates the presence of subcutaneous emphysema (SCE), pleural effusion (EFF), and pneumothorax (PNO). The localization of the boxes point to the region of interest of the deep learning algorithm, i.e. the region where a potentially pathological region is suspected.}}
\end{figure}

The proposed DLAD (Carebot AI CXR, \autoref{fig:showcase}) is a deep learning-based medical device designed to assist radiologists in interpreting chest X-ray images acquired in anteroposterior (AP) or posteroanterior (PA) projection. By employing advanced deep learning algorithms, this solution enables automatic detection of abnormal findings by analyzing visual patterns associated with specific conditions. The targeted abnormalities include atelectasis (ATE), consolidation (CON), pleural effusion (EFF), pulmonary lesion (LES), subcutaneous emphysema (SCE), cardiomegaly (CMG), and pneumothorax (PNO). The DLAD functions as a prediction algorithm complemented by various application peripherals, such as web-based communication tools, DICOM file conversion capabilities, and storage and reporting libraries supporting both DICOM Structured Report and DICOM Presentation State formats. 

\newpage

\subsection{Training Data}
A total of \fbox{\phantom{XX XXX}} chest X-rays with established ground truth from sites in Europe, Asia, and North America were used in the development of the DLAD. The complete training dataset includes \fbox{\phantom{XX XXX}} images with visually confirmed pathological findings, and \fbox{\phantom{XX XXX}} images with no or insignificant abnormal pathological findings. Chest radiographs were taken in posteroanterior (PA) or anteroposterior (AP) projection. As the images were obtained from the contacted centers in an anonymized form without additional metadata, the manufacturer of the DLAD does not have knowledge of the patients' history.

\section{Methodology}
The collected X-ray images were subjected to the proposed DLAD (Carebot AI CXR) for analysis. Subsequently, the DLAD's performance was compared with the standard clinical practice, where radiologists assessed the CXR images in the simulated hospital setting with access to standard viewing tools (e.g., pan, zoom, WW/WL) and were given an unlimited amount of time to complete the review. Each radiologist determined the presence or absence of 7 indicated radiological findings, including atelectasis (ATE), consolidation (CON), pleural effusion (EFF), pulmonary lesion (LES), subcutaneous emphysema (SCE), cardiomegaly (CMG), and pneumothorax (PNO).

\subsection{Data Source}
\begin{figure}[H]
\centering
\includegraphics[width=1\textwidth]{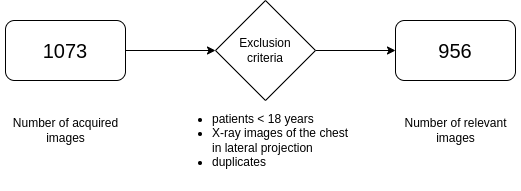}
\caption{\label{fig:method}\textbf\scriptsize{Methodology for selecting relevant chest X-ray images for the proposed study.}}
\end{figure}

In the period between October 18th, 2022, and November 17th, 2022, anonymized chest X-ray images of patients were collected at the Radiodiagnostic Department of the Havířov Hospital, p.o. The collection process involved utilizing the CloudPACS imaging and archiving system provided by OR-CZ spol. s r.o. A total of 1,073 chest X-rays were acquired within the specified period at the department. The data collection remained intact and unaffected throughout the testing phase, ensuring the integrity of the dataset. The collected sample accurately represents the prevalence of findings within the observed population. After excluding ineligible studies such as X-rays from patients under 18 years of age, lateral projection X-rays, and scans of insufficient quality (\autoref{fig:method}), a total of 956 relevant CXRs were identified for further assessment.
\newpage

\subsection{Ground Truth}
To confirm the presence or absence of a finding on the chest X-ray, a 100\% consensus of two highly experienced, board-certified radiologists was required to establish the ground truth (\autoref{tab:GT}). The agreement was observed at the level of individual indications, i.e., disagreement in the evaluation of one finding was not taken into account for the inclusion of an image in case of an agreement for another finding.

\begin{table}[H]
\centering
\adjustbox{max width=\textwidth}{
\begin{tabular}{|l|l|}
\hline
\textbf{Ground truth ID} & \textbf{Experience}                                                                                                                                           \\ \hline
\#c1235                        & \begin{tabular}[c]{@{}l@{}}Head of local radiology department, \textgreater{}20 years of experience, \\ board-certified\end{tabular}                          \\ \hline
\#24a8d                        & \begin{tabular}[c]{@{}l@{}}Head of the radiology department of a medium-sized hospital, \\ \textgreater{}20 years of experience, board-certified\end{tabular} \\ \hline
\end{tabular}}
\caption{\label{tab:GT} Ground truth annotators and the corresponding experience.}
\end{table}

\subsection{Objectives}
The primary objective is to evaluate the performance parameters of the proposed DLAD (Carebot AI CXR) in comparison to individual radiologists.

\subsection{Statistical Analysis}
The performance of DLAD was conducted using various statistical measures. These measures included sensitivity ensitivity \textit{(Se)} and specificity \textit{(Sp)}, positive \textit{(PLR)} and negative likelihood ratio \textit{(NLR)}, and positive \textit{(PPV)} and negative predictive value \textit{(NPV)}. Sensitivity represents the rate of true positive cases, while specificity represents the rate of false positive cases. The relationship between sensitivity and specificity is expressed by the formulas \textit{PLR = Se/(1-Sp)} and \textit{NLR = (1-Se)/Sp}. The likelihood ratios \textit{(LRs)} solely depend on sensitivity and specificity and are equivalent to the relative risk. It is desirable to have higher \textit{PLR} and lower \textit{NLR} values. Predictive values \textit{(PVs)} indicate the clinical accuracy of the diagnostic test and depend on sensitivity, specificity, and the prevalence of the disease in the population. In this study, a paired design was employed, where all images were evaluated by both the DLAD system and individual radiologists. The results were then compared against the ground truth. 

To address the primary objective, which involved comparing DLAD performance with individual radiologists, the aforementioned parameters were estimated and statistically compared using confidence intervals (CI) and \textit{p}-Values. The statistical comparison procedure consisted of two steps. Firstly, a global hypothesis test was conducted to determine whether there were significant differences between DLAD and radiologists in terms of both sensitivity and specificity (e.g. $H_0: (Se_1 = Se_2$ \textit{and} $Sp_1 = Sp_2)$ \textit{vs.} $H_1: (Se_1 \neq Se_2$ \textit{and/or} $Sp_1 \neq Sp_2)$). If the global hypothesis test yielded a significant result, individual hypothesis tests were performed. These tests compared sensitivity and specificity separately (e.g. $H_0: Se_1 = Se_2$ \textit{and} $H_0: Sp_1 = Sp_2$.

Additionally, a multiple comparison method, such as McNemar with continuity correction for \textit{Se} and \textit{Sp}, Holm method for \textit{LRs}, and weighted generalized score statistics for \textit{PVs}, was applied to control the overall error rate $\alpha$. Differences among radiologists and DLAD were graphically presented using forest plots. All tests were performed as two-tailed tests at the 5\% significance level.

\newpage
\subsection{Demographic Data and Prevalence of Individual Findings}
\begin{table}[H]
\centering
\begin{tabular}{|l|l|l|}
\hline
\textbf{Patient’s Sex} & \textbf{n} & \textbf{\%} \\ \hline
\quad Female                 & 480        & 50.21       \\ \hline
\quad Male                   & 474        & 49.58       \\ \hline
\quad Unspecified            & 2          & 0.21        \\ \hline
\textbf{Patient’s Age} & \textbf{}  & \textbf{}   \\ \hline
\quad 18-30                  & 58         & 6.07        \\ \hline
\quad 31-50                  & 163        & 17.05       \\ \hline
\quad 51-70                  & 366        & 38.28       \\ \hline
\quad 70+                    & 369        & 38.60       \\ \hline
\end{tabular}
\caption{\label{tab:DEM}Demographic data of the examined patients.}
\end{table}

\begin{table}[H]
\centering
\begin{tabular}{|l|c|}
\hline
\textbf{Finding}              & \textbf{n / N (prevalence)} \\ \hline
Atelectasis (ATE)                   & 6 / 908 (0.7\%)             \\ \hline
Consolidation (CON)                 & 22 / 830 (2.7\%)            \\ \hline
Pleural effusion (EFF)             & 85 / 909 (9.4\%)            \\ \hline
Lesion of the lung parenchyma (LES) & 21 / 901 (2.3\%)            \\ \hline
Subcutaneous emphysema (SCE)        & 2 / 953 (0.2\%)             \\ \hline
Cardiomegaly (CMG)                  & 49 / 865 (5.7\%)            \\ \hline
Pneumothorax (PNO)                  & 8 / 947 (0.8\%)             \\ \hline
\end{tabular}
\caption{\label{tab:PREVALENCE}The prevalence of observed individual findings.}
\end{table}

\begin{table}[]
\centering
\adjustbox{max width=\textwidth}{
\begin{tabular}{|l|l|}
\hline
\textbf{ID} & \textbf{Experience}                                         \\ \hline
\#5f049                       & Junior radiologist, \textless{}2 years of experience, without board-certification          \\ \hline
\#44247                       & Junior radiologist, \textless{}2 years of experience, without board-certification          \\ \hline
\#1c96c                       & Mid-level radiologist, \textless{}5 years of experience, without board-certification \\ \hline
\#e5ee5                       & Mid-level radiologist, \textless{}5 years of experience, without board-certification \\ \hline
\#cd16c                       & Experienced radiologist, \textgreater{}5 years of experience, board-certified        \\ \hline
\#b3ca6                       & Experienced radiologist, \textgreater{}5 years of experience, board-certified        \\ \hline
\end{tabular}}
\caption{\label{tab:RDG}List of radiologists involved in the multi-reader study and the corresponding experience.}
\end{table}

\newpage

\section{Results}
\begin{figure}[H]
\centering
\includegraphics[width=0.8\textwidth]{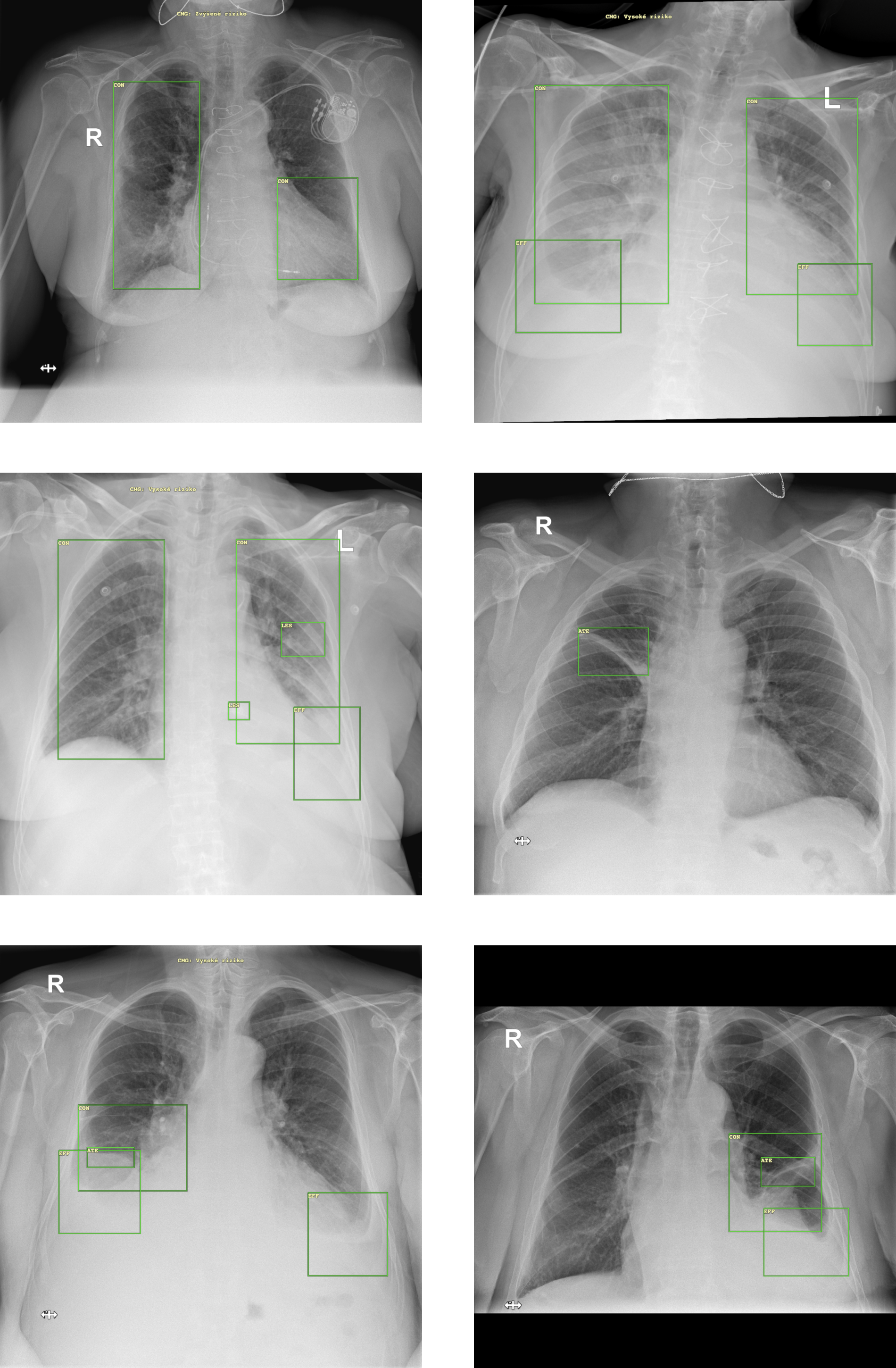}
\caption{\label{fig:examples}\textbf\scriptsize{Examples of predicted CXRs using the proposed DLAD (Carebot AI CXR).}}
\end{figure}
\newpage

\subsection{Atelectasis}
\begin{table}[H]
\centering
\adjustbox{max width=\textwidth}{
\begin{tabular}[width=1\textwidth]{|l|c|c|c|c|c|c|c|c|c|}
\hline
\textbf{ID}      & \textbf{Se}    & \textbf{LSe}   & \textbf{USe}   & \textbf{Sp}    & \textbf{LSp}   & \textbf{USp}   & \textbf{Global \textit{p}-Value}        & \textbf{Se \textit{p}-Value}           & \textbf{Sp \textit{p}-Value}           \\ \hline
\textbf{DLAD} & \textbf{1.000} & \textbf{0.624} & \textbf{1.000} & \textbf{0.905} & \textbf{0.884} & \textbf{0.922} & \textbf{}                     & \textbf{}                     & \textbf{}                     \\ \hline
\textbf{\#5f049} & 0.000          & 0.000          & 0.376          & 1.000          & 0.996          & 1.000          & 0.000 & 0.000 & 0.000 \\ \hline
\textbf{\#44247} & 0.833          & 0.446          & 0.979          & 0.968          & 0.954          & 0.978          & 0.000 & 0.230                         & 0.000 \\ \hline
\textbf{\#1c96c} & 0.833          & 0.446          & 0.979          & 0.906          & 0.885          & 0.923          & 0.546                         & 0.230                         & 0.993                         \\ \hline
\textbf{\#e5ee5} & 0.500          & 0.188          & 0.812          & 0.945          & 0.928          & 0.958          & 0.000 & 0.000 & 0.000 \\ \hline
\textbf{\#cd16c} & 1.000          & 0.624          & 1.000          & 0.759          & 0.731          & 0.786          & 1.000                         & 1.000                         & 0.000 \\ \hline
\textbf{\#b3ca6} & 0.500          & 0.188          & 0.812          & 0.947          & 0.930          & 0.960          & 0.000 & 0.000 & 0.000 \\ \hline
\textbf{ID}      & \textbf{PLR}   & \textbf{LPLR}  & \textbf{UPLR}  & \textbf{NLR}   & \textbf{LNLR}  & \textbf{UNLR}  & \textbf{Global \textit{p}-Value}        & \textbf{PLR \textit{p}-Value}          & \textbf{NLR \textit{p}-Value}          \\ \hline
\textbf{DLAD} & \textbf{10.49} & \textbf{5.68}  & \textbf{11.95} & \textbf{0.000} & \textbf{0.000} & \textbf{0.485} & \textbf{}                     &                               & \textbf{}                     \\ \hline
\textbf{\#5f049} &               &               &               & 1.000          & 0.561          & 0.993          &                              &                              &                              \\ \hline
\textbf{\#44247} & 25.92          & 12.13          & 35.83          & 0.172          & 0.052          & 0.600          &                              &                              &                              \\ \hline
\textbf{\#1c96c} & 8.84           & 4.35           & 10.51          & 0.184          & 0.056          & 0.641          &                              &                              &                              \\ \hline
\textbf{\#e5ee5} & 9.02           & 3.55           & 15.20          & 0.529          & 0.213          & 0.846          &                              &                              &                              \\ \hline
\textbf{\#cd16c} & 4.16           & 2.30           & 4.35           & 0.000          & 0.000          & 0.579          &                              &                              &                              \\ \hline
\textbf{\#b3ca6} & 9.40           & 3.69           & 15.87          & 0.528          & 0.213          & 0.844          &                              &                              &                              \\ \hline
\textbf{ID}      & \textbf{PPV}   & \textbf{LPPV}  & \textbf{UPPV}  & \textbf{NPV}   & \textbf{LNPV}  & \textbf{UNPV}  & \textbf{Global \textit{p}-Value}        & \textbf{PPV \textit{p}-Value}          & \textbf{NPV \textit{p}-Value}          \\ \hline
\textbf{DLAD} & \textbf{0.065} & \textbf{0.029} & \textbf{0.134} & \textbf{1.000} & \textbf{0.995} & \textbf{1.000} & \textbf{}                     & \textbf{}                     & \textbf{}                     \\ \hline
\textbf{\#5f049} &               &               &               & 0.993          & 0.986          & 0.997          &                               &                               &                               \\ \hline
\textbf{\#44247} & 0.147          & 0.062          & 0.299          & 0.999          & 0.994          & 1.000          & 0.043 & 0.000 & 0.334                         \\ \hline
\textbf{\#1c96c} & 0.056          & 0.023          & 0.122          & 0.999          & 0.993          & 1.000          & 0.604                         & 0.436                         & 0.318                         \\ \hline
\textbf{\#e5ee5} & 0.057          & 0.017          & 0.152          & 0.996          & 0.990          & 0.999          & 0.080                         & 0.728                         & 0.090                         \\ \hline
\textbf{\#cd16c} & 0.027          & 0.012          & 0.057          & 1.000          & 0.995          & 1.000          &                               &                               &                               \\ \hline
\textbf{\#b3ca6} & 0.059          & 0.018          & 0.157          & 0.996          & 0.990          & 0.999          & 0.077                         & 0.800                         & 0.090                         \\ \hline
\end{tabular}}
\caption{\label{tab:ATE}Performance of the proposed DLAD and assessed radiologists for the finding atelectasis (ATE).}
\end{table}
The atelectasis (ATE) has a low prevalence of a confirmed finding (prevalence: 0.7\%). The DLAD identified all positive cases as ATE+, achieving sensitivity (\textit{Se}) of 1.000 (0.624-1.000). Additionally, the DLAD predicted 86 CXRs as ATE+ that were actually ATE-, resulting in specificity (\textit{Sp}) of 0.905 (0.884-0.922). Considering the low prevalence, the reliability of the results is limited. However, the DLAD still outperformed the radiologists, although some radiologists exhibited higher \textit{Sp} at the expense of lower \textit{Se}. Due to the rarity of the diagnoses, a comparison of likelihood ratios was not possible, and certain statistical measures could not be estimated.

\newpage

\subsection{Consolidation}
\begin{table}[H]
\centering
\adjustbox{max width=\textwidth}{
\begin{tabular}[width=1\textwidth]{|l|c|c|c|c|c|c|c|c|c|}
\hline
\textbf{ID}      & \textbf{Se}    & \textbf{LSe}   & \textbf{USe}   & \textbf{Sp}    & \textbf{LSp}   & \textbf{USp}   & \textbf{Global \textit{p}-Value}        & \textbf{Se \textit{p}-Value}           & \textbf{Sp \textit{p}-Value}           \\ \hline
\textbf{DLAD}    & \textbf{0.864} & \textbf{0.671} & \textbf{0.956} & \textbf{0.854} & \textbf{0.828} & \textbf{0.877} & \textbf{}                     & \textbf{}                     & \textbf{}                     \\ \hline
\textbf{\#5f049} & 0.182          & 0.070          & 0.382          & 0.998          & 0.991          & 0.999          & 0.000 & 0.000 & 0.000 \\ \hline
\textbf{\#44247} & 0.545          & 0.347          & 0.731          & 0.868          & 0.843          & 0.889          & 0.019 & 0.000 & 0.468                         \\ \hline
\textbf{\#1c96c} & 0.545          & 0.347          & 0.731          & 0.990          & 0.981          & 0.995          & 0.000 & 0.000 & 0.000 \\ \hline
\textbf{\#e5ee5} & 0.864          & 0.671          & 0.956          & 0.979          & 0.967          & 0.987          & 0.000 & 1.000                         & 0.000 \\ \hline
\textbf{\#cd16c} & 0.909          & 0.726          & 0.979          & 0.855          & 0.829          & 0.878          & 0.590                         & 0.295                         & 0.994                         \\ \hline
\textbf{\#b3ca6} & 0.500          & 0.307          & 0.693          & 0.975          & 0.962          & 0.984          & 0.000 & 0.000 & 0.000 \\ \hline
\textbf{ID}      & \textbf{PLR}   & \textbf{LPLR}  & \textbf{UPLR}  & \textbf{NLR}   & \textbf{LNLR}  & \textbf{UNLR}  & \textbf{Global \textit{p}-Value}        & \textbf{PLR \textit{p}-Value}          & \textbf{NLR \textit{p}-Value}          \\ \hline
\textbf{DLAD}    & \textbf{5.91}  & \textbf{4.26}  & \textbf{6.93}  & \textbf{0.160} & \textbf{0.066} & \textbf{0.406} & \textbf{}                     & \textbf{}                     & \textbf{}                     \\ \hline
\textbf{\#5f049} & 73.45          & 20.80          & 567.33         & 0.820          & 0.607          & 0.921          & 0.000 & 0.002 & 0.002 \\ \hline
\textbf{\#44247} & 4.12           & 2.56           & 5.75           & 0.524          & 0.315          & 0.753          & 0.088                         & 0.115                         & 0.030 \\ \hline
\textbf{\#1c96c} & 55.09          & 26.57          & 124.66         & 0.459          & 0.277          & 0.659          & 0.000 & 0.000 & 0.054                         \\ \hline
\textbf{\#e5ee5} & 41.05          & 24.40          & 64.60          & 0.139          & 0.058          & 0.354          & 0.000 & 0.000 & 0.838                         \\ \hline
\textbf{\#cd16c} & 6.28           & 4.61           & 7.33           & 0.106          & 0.039          & 0.344          & 0.603                         & 0.600                         & 0.319                         \\ \hline
\textbf{\#b3ca6} & 20.20          & 11.03          & 35.69          & 0.513          & 0.318          & 0.708          & 0.000 & 0.000 & 0.053                         \\ \hline
\textbf{ID}      & \textbf{PPV}   & \textbf{LPPV}  & \textbf{UPPV}  & \textbf{NPV}   & \textbf{LNPV}  & \textbf{UNPV}  & \textbf{Global \textit{p}-Value}        & \textbf{PPV \textit{p}-Value}          & \textbf{NPV \textit{p}-Value}          \\ \hline
\textbf{DLAD}    & \textbf{0.139} & \textbf{0.090} & \textbf{0.206} & \textbf{0.996} & \textbf{0.988} & \textbf{0.999} & \textbf{}                     & \textbf{}                     & \textbf{}                     \\ \hline
\textbf{\#5f049} & 0.667          & 0.305          & 0.908          & 0.978          & 0.966          & 0.986          & 0.000 & 0.000 & 0.001 \\ \hline
\textbf{\#44247} & 0.101          & 0.058          & 0.167          & 0.986          & 0.974          & 0.993          & 0.055                         & 0.115                         & 0.022 \\ \hline
\textbf{\#1c96c} & 0.600          & 0.388          & 0.782          & 0.988          & 0.978          & 0.993          & 0.000 & 0.000 & 0.044 \\ \hline
\textbf{\#e5ee5} & 0.528          & 0.370          & 0.680          & 0.996          & 0.989          & 0.999          & 0.000 & 0.000 & 0.838                         \\ \hline
\textbf{\#cd16c} & 0.146          & 0.096          & 0.214          & 0.997          & 0.990          & 0.999          & 0.604                         & 0.600                         & 0.316                         \\ \hline
\textbf{\#b3ca6} & 0.355          & 0.210          & 0.529          & 0.986          & 0.976          & 0.992          & 0.000 & 0.000 & 0.041 \\ \hline
\end{tabular}}
\caption{\label{tab:CON}Performance of the proposed DLAD and assessed radiologists for the finding consolidation (CON).}
\end{table}

A total of 22 images (prevalence: 2.7\%) were with confirmed consolidation (CON). The DLAD successfully identified 19 of these as CON+, resulting in \textit{Se} of 0.864 (0.671-0.956). Additionally, the DLAD incorrectly flagged 118 images as CON+ that were, in fact, CON-, indicating \textit{Sp} of 0.854 (0.828-0.877). Notably, only two radiologists, who possessed more experience, achieved a similar balance between \textit{Se} and \textit{Sp} in this diagnosis. The increased number of false positive scans can be attributed to the lowest agreement among the physicians involved in determining the ground truth for this particular finding.

\newpage

\subsection{Pleural Effusion}
\begin{table}[H]
\centering
\adjustbox{max width=\textwidth}{
\begin{tabular}[width=1\textwidth]{|l|c|c|c|c|c|c|c|c|c|}
\hline
\textbf{ID}      & \textbf{Se}    & \textbf{LSe}   & \textbf{USe}   & \textbf{Sp}    & \textbf{LSp}   & \textbf{USp}   & \textbf{Global \textit{p}-Value}        & \textbf{Se \textit{p}-Value}           & \textbf{Sp \textit{p}-Value}           \\ \hline
\textbf{DLAD}    & \textbf{0.953} & \textbf{0.887} & \textbf{0.983} & \textbf{0.876} & \textbf{0.852} & \textbf{0.897} & \textbf{}                     & \textbf{}                     & \textbf{}                     \\ \hline
\textbf{\#5f049} & 0.400          & 0.302          & 0.506          & 1.000          & 0.996          & 1.000          & 0.000 & 0.000 & 0.000 \\ \hline
\textbf{\#44247} & 0.576          & 0.471          & 0.676          & 0.975          & 0.962          & 0.983          & 0.000 & 0.000 & 0.000 \\ \hline
\textbf{\#1c96c} & 0.753          & 0.652          & 0.833          & 0.989          & 0.980          & 0.994          & 0.000 & 0.000 & 0.000 \\ \hline
\textbf{\#e5ee5} & 0.953          & 0.887          & 0.983          & 0.982          & 0.970          & 0.989          & 0.000 & 1.000                         & 0.000 \\ \hline
\textbf{\#cd16c} & 0.882          & 0.798          & 0.936          & 0.966          & 0.951          & 0.977          & 0.000 & 0.002 & 0.000 \\ \hline
\textbf{\#b3ca6} & 0.753          & 0.652          & 0.833          & 0.990          & 0.981          & 0.995          & 0.000 & 0.000 & 0.000 \\ \hline
\textbf{ID}      & \textbf{PLR}   & \textbf{LPLR}  & \textbf{UPLR}  & \textbf{NLR}   & \textbf{LNLR}  & \textbf{UNLR}  & \textbf{Global \textit{p}-Value}        & \textbf{PLR \textit{p}-Value}          & \textbf{NLR \textit{p}-Value}          \\ \hline
\textbf{DLAD}    & \textbf{7.70}  & \textbf{6.29}  & \textbf{9.20}  & \textbf{0.054} & \textbf{0.023} & \textbf{0.136} & \textbf{}                     & \textbf{}                     & \textbf{}                     \\ \hline
\textbf{\#5f049} &                &                &                & 0.600          & 0.493          & 0.697          &                               &                               &                               \\ \hline
\textbf{\#44247} & 22.62          & 14.67          & 37.58          & 0.435          & 0.333          & 0.544          & 0.000 & 0.000 & 0.000 \\ \hline
\textbf{\#1c96c} & 68.94          & 38.44          & 133.29         & 0.250          & 0.172          & 0.354          & 0.000 & 0.000 & 0.001 \\ \hline
\textbf{\#e5ee5} & 52.35          & 32.95          & 83.41          & 0.048          & 0.022          & 0.122          & 0.000 & 0.000 & 0.872                         \\ \hline
\textbf{\#cd16c} & 25.97          & 18.14          & 37.32          & 0.122          & 0.070          & 0.214          & 0.000 & 0.000 & 0.135                         \\ \hline
\textbf{\#b3ca6} & 77.55          & 42.01          & 154.98         & 0.249          & 0.172          & 0.354          & 0.000 & 0.000 & 0.001 \\ \hline
\textbf{ID}      & \textbf{PPV}   & \textbf{LPPV}  & \textbf{UPPV}  & \textbf{NPV}   & \textbf{LNPV}  & \textbf{UNPV}  & \textbf{Global \textit{p}-Value}        & \textbf{PPV \textit{p}-Value}          & \textbf{NPV \textit{p}-Value}          \\ \hline
\textbf{DLAD}    & \textbf{0.443} & \textbf{0.372} & \textbf{0.515} & \textbf{0.994} & \textbf{0.986} & \textbf{0.998} & \textbf{}                     & \textbf{}                     & \textbf{}                     \\ \hline
\textbf{\#5f049} & 1.000          & 0.902          & 1.000          & 0.942          & 0.924          & 0.956          & 0.000 & 0.000 & 0.000 \\ \hline
\textbf{\#44247} & 0.700          & 0.585          & 0.795          & 0.957          & 0.941          & 0.969          & 0.000 & 0.000 & 0.000 \\ \hline
\textbf{\#1c96c} & 0.877          & 0.783          & 0.935          & 0.975          & 0.962          & 0.984          & 0.000 & 0.000 & 0.000 \\ \hline
\textbf{\#e5ee5} & 0.844          & 0.759          & 0.904          & 0.995          & 0.988          & 0.998          & 0.000 & 0.000 & 0.872                         \\ \hline
\textbf{\#cd16c} & 0.728          & 0.636          & 0.805          & 0.988          & 0.977          & 0.993          & 0.000 & 0.000 & 0.124                         \\ \hline
\textbf{\#b3ca6} & 0.889          & 0.797          & 0.944          & 0.975          & 0.962          & 0.984          & 0.000 & 0.000 & 0.000 \\ \hline
\end{tabular}}
\caption{\label{tab:EFF}Performance of the proposed DLAD and assessed radiologists for the finding pleural effusion (EFF).}
\end{table}

A total of 85 images (prevalence: 9.4\%) exhibited pleural effusion (EFF). The DLAD accurately identified 81 of these as EFF+, resulting in \textit{Se} of 0.953 (0.887-0.983). However, the DLAD also incorrectly labeled 102 images as EFF+ that were, in fact, EFF-, leading to \textit{Sp} of 0.876 (0.852-0.897). Notably, only two more experienced radiologists achieved a similar balance between \textit{Se} and \textit{Sp} in this diagnosis. It is important to highlight that the DLAD achieved a lower positive predictive value (\textit{PPV}) due to the higher number of false positive images.

\newpage

\subsection{Pulmonary Lesion}

\begin{table}[H]
\centering
\adjustbox{max width=\textwidth}{
\begin{tabular}[width=1\textwidth]{|l|c|c|c|c|c|c|c|c|c|}
\hline
\textbf{ID}      & \textbf{Se}    & \textbf{LSe}   & \textbf{USe}   & \textbf{Sp}    & \textbf{LSp}   & \textbf{USp}   & \textbf{Global \textit{p}-Value}        & \textbf{Se \textit{p}-Value}           & \textbf{Sp \textit{p}-Value}           \\ \hline
\textbf{DLAD}    & \textbf{0.905} & \textbf{0.715} & \textbf{0.978} & \textbf{0.893} & \textbf{0.871} & \textbf{0.912} & \textbf{}                     & \textbf{}                     & \textbf{}                     \\ \hline
\textbf{\#5f049} & 0.238          & 0.103          & 0.448          & 0.999          & 0.994          & 1.000          & 0.000 & 0.000 & 0.000 \\ \hline
\textbf{\#44247} & 0.333          & 0.170          & 0.544          & 0.933          & 0.915          & 0.948          & 0.000 & 0.000 & 0.000 \\ \hline
\textbf{\#1c96c} & 0.524          & 0.324          & 0.717          & 0.884          & 0.861          & 0.904          & 0.001 & 0.000 & 0.685                         \\ \hline
\textbf{\#e5ee5} & 0.619          & 0.410          & 0.794          & 0.968          & 0.955          & 0.978          & 0.000 & 0.000 & 0.000 \\ \hline
\textbf{\#cd16c} & 0.667          & 0.456          & 0.830          & 0.991          & 0.982          & 0.996          & 0.000 & 0.000 & 0.000 \\ \hline
\textbf{\#b3ca6} & 0.619          & 0.410          & 0.794          & 0.989          & 0.979          & 0.994          & 0.000 & 0.000 & 0.000 \\ \hline
\textbf{ID}      & \textbf{PLR}   & \textbf{LPLR}  & \textbf{UPLR}  & \textbf{NLR}   & \textbf{LNLR}  & \textbf{UNLR}  & \textbf{Global \textit{p}-Value}        & \textbf{PLR \textit{p}-Value}          & \textbf{NLR \textit{p}-Value}          \\ \hline
\textbf{DLAD}    & \textbf{8.47}  & \textbf{6.08}  & \textbf{10.12} & \textbf{0.107} & \textbf{0.039} & \textbf{0.342} & \textbf{}                     & \textbf{}                     & \textbf{}                     \\ \hline
\textbf{\#5f049} & 209.52         & 45.83          & 348.21         & 0.763          & 0.543          & 0.887          & 0.000 & 0.003 & 0.003 \\ \hline
\textbf{\#44247} & 4.97           & 2.59           & 8.77           & 0.715          & 0.483          & 0.881          & 0.026 & 0.137                         & 0.007 \\ \hline
\textbf{\#1c96c} & 4.52           & 2.74           & 6.45           & 0.539          & 0.325          & 0.763          & 0.005 & 0.007 & 0.010 \\ \hline
\textbf{\#e5ee5} & 19.46          & 11.45          & 30.04          & 0.393          & 0.221          & 0.612          & 0.000 & 0.001 & 0.033 \\ \hline
\textbf{\#cd16c} & 73.33          & 36.22          & 152.57         & 0.336          & 0.182          & 0.554          & 0.000 & 0.000 & 0.055                         \\ \hline
\textbf{\#b3ca6} & 54.48          & 27.81          & 109.09         & 0.385          & 0.217          & 0.599          & 0.000 & 0.000 & 0.069                         \\ \hline
\textbf{ID}      & \textbf{PPV}   & \textbf{LPPV}  & \textbf{UPPV}  & \textbf{NPV}   & \textbf{LNPV}  & \textbf{UNPV}  & \textbf{Global \textit{p}-Value}        & \textbf{PPV \textit{p}-Value}          & \textbf{NPV \textit{p}-Value}          \\ \hline
\textbf{DLAD}    & \textbf{0.168} & \textbf{0.110} & \textbf{0.247} & \textbf{0.997} & \textbf{0.991} & \textbf{0.999} & \textbf{}                     & \textbf{}                     & \textbf{}                     \\ \hline
\textbf{\#5f049} & 0.833          & 0.446          & 0.979          & 0.982          & 0.971          & 0.989          & 0.000 & 0.000 & 0.001 \\ \hline
\textbf{\#44247} & 0.106          & 0.051          & 0.202          & 0.983          & 0.972          & 0.990          & 0.002 & 0.136                         & 0.002 \\ \hline
\textbf{\#1c96c} & 0.097          & 0.054          & 0.165          & 0.987          & 0.977          & 0.993          & 0.014 & 0.007 & 0.004 \\ \hline
\textbf{\#e5ee5} & 0.317          & 0.195          & 0.469          & 0.991          & 0.982          & 0.995          & 0.000 & 0.000 & 0.022 \\ \hline
\textbf{\#cd16c} & 0.636          & 0.431          & 0.804          & 0.992          & 0.984          & 0.996          & 0.000 & 0.000 & 0.042 \\ \hline
\textbf{\#b3ca6} & 0.565          & 0.369          & 0.744          & 0.991          & 0.982          & 0.996          & 0.000 & 0.000 & 0.052                         \\ \hline
\end{tabular}}
\caption{\label{tab:LES}Performance of the proposed DLAD and assessed radiologists for the finding pulmonary lesion (LES).}
\end{table}

A total of 21 scans (prevalence: 2.3\%) were with confirmed pulmonary lesions (LES). The DLAD correctly identified 19 of these as LES+, resulting in \textit{Se} of 0.905 (0.715-0.978). Additionally, the DLAD incorrectly flagged 94 images as LES+ that were actually LES-, leading to \textit{Sp} of 0.893 (0.871-0.912). Notably, none of the radiologists achieved a similar level of performance in this diagnosis. It is important to note that the low \textit{PPV} can be attributed to both a higher false positive rate and the low prevalence of pulmonary lesions in the dataset.

\newpage

\subsection{Subcutaneous Emphysema}
\begin{table}[]
\centering
\adjustbox{max width=\textwidth}{
\begin{tabular}[width=1\textwidth]{|l|c|c|c|c|c|c|c|c|c|}
\hline
\textbf{ID}      & \textbf{Se}    & \textbf{LSe}   & \textbf{USe}   & \textbf{Sp}    & \textbf{LSp}   & \textbf{USp}   & \textbf{Global \textit{p}-Value}        & \textbf{Se \textit{p}-Value}           & \textbf{Sp \textit{p}-Value}           \\ \hline
\textbf{DLAD}    & \textbf{1.000} & \textbf{0.366} & \textbf{1.000} & \textbf{0.966} & \textbf{0.953} & \textbf{0.976} & \textbf{}                     & \textbf{}                     & \textbf{}                     \\ \hline
\textbf{\#5f049} & 0.000          & 0.000          & 0.634          & 1.000          & 0.996          & 1.000          & 0.000 & 0.000 & 0.000 \\ \hline
\textbf{\#44247} & 0.500          & 0.095          & 0.905          & 1.000          & 0.996          & 1.000          & 0.000 & 0.046 & 0.000 \\ \hline
\textbf{\#1c96c} & 1.000          & 0.366          & 1.000          & 0.999          & 0.994          & 1.000          & 1.000                         & 1.000                         & 0.000 \\ \hline
\textbf{\#e5ee5} & 1.000          & 0.366          & 1.000          & 1.000          & 0.996          & 1.000          & 1.000                         & 1.000                         & 0.000 \\ \hline
\textbf{\#cd16c} & 1.000          & 0.366          & 1.000          & 1.000          & 0.996          & 1.000          & 1.000                         & 1.000                         & 0.000 \\ \hline
\textbf{\#b3ca6} & 1.000          & 0.366          & 1.000          & 1.000          & 0.996          & 1.000          & 1.000                         & 1.000                         & 0.000 \\ \hline
\textbf{ID}      & \textbf{PLR}   & \textbf{LPLR}  & \textbf{UPLR}  & \textbf{NLR}   & \textbf{LNLR}  & \textbf{UNLR}  & \textbf{Global \textit{p}-Value}        & \textbf{PLR \textit{p}-Value}          & \textbf{NLR \textit{p}-Value}          \\ \hline
\textbf{DLAD}    & \textbf{29.72} & \textbf{8.87}  & \textbf{35.31} & \textbf{0.000} & \textbf{0.000} & \textbf{0.715} & \textbf{}                     & \textbf{}                     & \textbf{}                     \\ \hline
\textbf{\#5f049} &                &                &                & 1.000          & 0.310          & 0.983          &                               &                               &                               \\ \hline
\textbf{\#44247} &                &                &                & 0.500          & 0.125          & 0.876          &                               &                               &                               \\ \hline
\textbf{\#1c96c} & 951.00         & 145.88         & 6902.31        & 0.000          & 0.000          & 0.692          &                               &                               &                               \\ \hline
\textbf{\#e5ee5} &                &                &                & 0.000          & 0.000          & 0.691          &                               &                               &                               \\ \hline
\textbf{\#cd16c} &                &                &                & 0.000          & 0.000          & 0.691          &                               &                               &                               \\ \hline
\textbf{\#b3ca6} &                &                &                & 0.000          & 0.000          & 0.691          &                               &                               &                               \\ \hline
\textbf{ID}      & \textbf{PPV}   & \textbf{LPPV}  & \textbf{UPPV}  & \textbf{NPV}   & \textbf{LNPV}  & \textbf{UNPV}  & \textbf{Global \textit{p}-Value}        & \textbf{PPV \textit{p}-Value}          & \textbf{NPV \textit{p}-Value}          \\ \hline
\textbf{DLAD}    & \textbf{0.059} & \textbf{0.013} & \textbf{0.188} & \textbf{1.000} & \textbf{0.996} & \textbf{1.000} & \textbf{}                     & \textbf{}                     & \textbf{}                     \\ \hline
\textbf{\#5f049} &                &                &                & 0.998          & 0.993          & 1.000          &                               &                               &                               \\ \hline
\textbf{\#44247} & 1.000          & 0.235          & 1.000          & 0.999          & 0.994          & 1.000          & 0.000 & 0.000 & 0.326                         \\ \hline
\textbf{\#1c96c} & 0.667          & 0.214          & 0.945          & 1.000          & 0.996          & 1.000          &                               &                               &                               \\ \hline
\textbf{\#e5ee5} & 1.000          & 0.366          & 1.000          & 1.000          & 0.996          & 1.000          &                               &                               &                               \\ \hline
\textbf{\#cd16c} & 1.000          & 0.366          & 1.000          & 1.000          & 0.996          & 1.000          &                               &                               &                               \\ \hline
\textbf{\#b3ca6} & 1.000          & 0.366          & 1.000          & 1.000          & 0.996          & 1.000          &                               &                               &                               \\ \hline
\end{tabular}}
\caption{\label{tab:SCE}Performance of the proposed DLAD and assessed radiologists for the finding subcutaneous emphysema (SCE).}
\end{table}
Subcutaneous emphysema (SCE) is an exceptionally rare condition, with only 2 scans (prevalence: 0.2\%) confirming its presence. The DLAD accurately identified both of these cases as SCE+, demonstrating \textit{Se} of 1.000 (0.366-1.000). However, given the small sample size, the confidence interval for \textit{Se} is wide. Additionally, the DLAD flagged an additional 32 images as SCE+, which were actually SCE-, resulting in \textit{Sp} of 0.966 (0.953-0.976). Generally, the assessment of this diagnosis was more effectively performed by assessed radiologists, as the DLAD exhibited a higher number of false positive predictions. Notably, less experienced radiologists demonstrated low detection rates, with \#5f049 not identifying any positive cases and \#44247 identifying only one. In such cases, the DLAD would have provided significant assistance to the junior radiologists.

\newpage

\subsection{Cardiomegaly}
\begin{table}[]
\centering
\adjustbox{max width=\textwidth}{
\begin{tabular}[width=1\textwidth]{|l|c|c|c|c|c|c|c|c|c|}
\hline
\textbf{ID}      & \textbf{Se}    & \textbf{LSe}   & \textbf{USe}   & \textbf{Sp}    & \textbf{LSp}   & \textbf{USp}   & \textbf{Global \textit{p}-Value}        & \textbf{Se \textit{p}-Value}           & \textbf{Sp \textit{p}-Value}           \\ \hline
\textbf{DLAD}    & \textbf{0.837} & \textbf{0.711} & \textbf{0.917} & \textbf{0.953} & \textbf{0.937} & \textbf{0.966} & \textbf{}                     & \textbf{}                     & \textbf{}                     \\ \hline
\textbf{\#5f049} & 0.347          & 0.228          & 0.486          & 0.990          & 0.981          & 0.995          & 0.000 & 0.000 & 0.000 \\ \hline
\textbf{\#44247} & 0.857          & 0.735          & 0.931          & 0.928          & 0.908          & 0.944          & 0.031 & 0.927                         & 0.000 \\ \hline
\textbf{\#1c96c} & 0.980          & 0.896          & 0.999          & 0.887          & 0.864          & 0.907          & 0.000 & 0.000 & 0.000 \\ \hline
\textbf{\#e5ee5} & 0.735          & 0.599          & 0.839          & 0.982          & 0.970          & 0.989          & 0.000 & 0.045 & 0.000 \\ \hline
\textbf{\#cd16c} & 0.796          & 0.665          & 0.887          & 0.978          & 0.966          & 0.986          & 0.005 & 0.737                         & 0.000 \\ \hline
\textbf{\#b3ca6} & 0.980          & 0.896          & 0.999          & 0.890          & 0.866          & 0.910          & 0.000 & 0.000 & 0.000 \\ \hline
\textbf{ID}      & \textbf{PLR}   & \textbf{LPLR}  & \textbf{UPLR}  & \textbf{NLR}   & \textbf{LNLR}  & \textbf{UNLR}  & \textbf{Global \textit{p}-Value}        & \textbf{PLR \textit{p}-Value}          & \textbf{NLR \textit{p}-Value}          \\ \hline
\textbf{DLAD}    & \textbf{17.97} & \textbf{12.72} & \textbf{24.70} & \textbf{0.171} & \textbf{0.094} & \textbf{0.310} & \textbf{}                     & \textbf{}                     & \textbf{}                     \\ \hline
\textbf{\#5f049} & 35.39          & 17.62          & 98.10          & 0.660          & 0.517          & 0.776          & 0.000 & 0.073                         & 0.000 \\ \hline
\textbf{\#44247} & 11.85          & 8.84           & 15.22          & 0.154          & 0.081          & 0.294          & 0.034 & 0.027 & 0.811                         \\ \hline
\textbf{\#1c96c} & 8.69           & 6.91           & 10.43          & 0.023          & 0.007          & 0.133          & 0.000 & 0.000 & 0.032 \\ \hline
\textbf{\#e5ee5} & 39.97          & 24.29          & 68.08          & 0.270          & 0.169          & 0.414          & 0.000 & 0.003 & 0.197                         \\ \hline
\textbf{\#cd16c} & 36.08          & 22.81          & 57.69          & 0.209          & 0.122          & 0.349          & 0.007 & 0.007 & 0.610                         \\ \hline
\textbf{\#b3ca6} & 8.88           & 7.04           & 10.68          & 0.023          & 0.007          & 0.132          & 0.000 & 0.000 & 0.058                         \\ \hline
\textbf{ID}      & \textbf{PPV}   & \textbf{LPPV}  & \textbf{UPPV}  & \textbf{NPV}   & \textbf{LNPV}  & \textbf{UNPV}  & \textbf{Global \textit{p}-Value}        & \textbf{PPV \textit{p}-Value}          & \textbf{NPV \textit{p}-Value}          \\ \hline
\textbf{DLAD}    & \textbf{0.519} & \textbf{0.411} & \textbf{0.626} & \textbf{0.990} & \textbf{0.980} & \textbf{0.995} & \textbf{}                     & \textbf{}                     & \textbf{}                     \\ \hline
\textbf{\#5f049} & 0.680          & 0.486          & 0.830          & 0.962          & 0.947          & 0.973          & 0.000 & 0.065                         & 0.000 \\ \hline
\textbf{\#44247} & 0.416          & 0.324          & 0.513          & 0.991          & 0.981          & 0.996          & 0.034 & 0.026 & 0.811                         \\ \hline
\textbf{\#1c96c} & 0.343          & 0.269          & 0.424          & 0.999          & 0.992          & 1.000          & 0.000 & 0.000 & 0.012 \\ \hline
\textbf{\#e5ee5} & 0.706          & 0.571          & 0.814          & 0.984          & 0.973          & 0.991          & 0.000 & 0.002 & 0.193                         \\ \hline
\textbf{\#cd16c} & 0.684          & 0.556          & 0.791          & 0.988          & 0.978          & 0.993          & 0.004 & 0.005 & 0.609                         \\ \hline
\textbf{\#b3ca6} & 0.348          & 0.273          & 0.430          & 0.999          & 0.992          & 1.000          & 0.000 & 0.000 & 0.026 \\ \hline
\end{tabular}}
\caption{\label{tab:CMG}Performance of the proposed DLAD and assessed radiologists for the finding atelectasis cardiomegaly (CMG).}
\end{table}

A total of 49 scans (prevalence: 5.7\%) were confirmed to have cardiomegaly (CMG). The DLAD accurately identified 41 of these cases as abnormal, resulting in \textit{Se} of 0.837 (0.711-0.917). Additionally, the DLAD assessed 38 images as CMG+ that were actually CMG-, leading to \textit{Sp} of 0.953 (0.937-0.966). Notably, in this diagnosis, the DLAD demonstrated a high level of \textit{Sp} and maintained an acceptable level of \textit{Se}. Although three radiologists achieved higher \textit{Se}, their \textit{Sp} was lower. Other characteristics of the DLAD exhibited similar trends. Given the DLAD's supportive purpose, the very good \textit{PPV} suggests that implementing the DLAD into the clinical workflow can provide benefits without imposing any additional workload on the radiologist.

\newpage

\subsection{Pneumothorax}

\begin{table}[H]
\centering
\adjustbox{max width=\textwidth}{
\begin{tabular}{|l|c|c|c|c|c|c|c|c|c|}
\hline
\textbf{ID}      & \textbf{Se}    & \textbf{LSe}   & \textbf{USe}   & \textbf{Sp}    & \textbf{LSp}   & \textbf{USp}   & \textbf{Global \textit{p}-Value}        & \textbf{Se \textit{p}-Value}           & \textbf{Sp \textit{p}-Value}           \\ \hline
\textbf{DLAD}    & \textbf{0.875} & \textbf{0.538} & \textbf{0.986} & \textbf{0.922} & \textbf{0.903} & \textbf{0.938} & \textbf{}                     & \textbf{}                     & \textbf{}                     \\ \hline
\textbf{\#5f049} & 0.500          & 0.215          & 0.785          & 1.000          & 0.996          & 1.000          & 0.000 & 0.000 & 0.000 \\ \hline
\textbf{\#44247} & 0.375          & 0.134          & 0.691          & 0.986          & 0.977          & 0.992          & 0.000 & 0.000 & 0.000 \\ \hline
\textbf{\#1c96c} & 0.750          & 0.415          & 0.934          & 0.996          & 0.989          & 0.998          & 0.000 & 0.253                         & 0.000 \\ \hline
\textbf{\#e5ee5} & 0.750          & 0.415          & 0.934          & 0.997          & 0.991          & 0.999          & 0.000 & 0.253                         & 0.000 \\ \hline
\textbf{\#cd16c} & 0.750          & 0.415          & 0.934          & 0.999          & 0.994          & 1.000          & 0.000 & 0.253                         & 0.000 \\ \hline
\textbf{\#b3ca6} & 0.875          & 0.538          & 0.986          & 0.999          & 0.994          & 1.000          & 1.000                         & 1.000                         & 0.000 \\ \hline
\textbf{ID}      & \textbf{PLR}   & \textbf{LPLR}  & \textbf{UPLR}  & \textbf{NLR}   & \textbf{LNLR}  & \textbf{UNLR}  & \textbf{Global \textit{p}-Value}        & \textbf{PLR \textit{p}-Value}          & \textbf{NLR \textit{p}-Value}          \\ \hline
\textbf{DLAD}    & \textbf{11.26} & \textbf{6.26}  & \textbf{13.55} & \textbf{0.136} & \textbf{0.042} & \textbf{0.535} & \textbf{}                     & \textbf{}                     & \textbf{}                     \\ \hline
\textbf{\#5f049} &                &                &                & 0.500          & 0.227          & 0.775          &                               &                               &                               \\ \hline
\textbf{\#44247} & 27.09          & 10.09          & 63.05          & 0.634          & 0.313          & 0.860          & 0.022 & 0.095                         & 0.085                         \\ \hline
\textbf{\#1c96c} & 176.06         & 64.11          & 426.23         & 0.251          & 0.091          & 0.601          & 0.000 & 0.000 & 0.383                         \\ \hline
\textbf{\#e5ee5} & 234.75         & 78.68          & 621.72         & 0.251          & 0.091          & 0.601          & 0.000 & 0.000 & 0.384                         \\ \hline
\textbf{\#cd16c} & 704.25         & 151.33         & 1208.39        & 0.250          & 0.090          & 0.599          & 0.000 & 0.000 & 0.386                         \\ \hline
\textbf{\#b3ca6} & 821.62         & 178.89         & 1361.98        & 0.125          & 0.039          & 0.494          & 0.000 & 0.000 & 0.000 \\ \hline
\textbf{ID}      & \textbf{PPV}   & \textbf{LPPV}  & \textbf{UPPV}  & \textbf{NPV}   & \textbf{LNPV}  & \textbf{UNPV}  & \textbf{Global \textit{p}-Value}        & \textbf{PPV \textit{p}-Value}          & \textbf{NPV \textit{p}-Value}          \\ \hline
\textbf{DLAD}    & \textbf{0.087} & \textbf{0.042} & \textbf{0.168} & \textbf{0.999} & \textbf{0.994} & \textbf{1.000} & \textbf{}                     & \textbf{}                     & \textbf{}                     \\ \hline
\textbf{\#5f049} & 1.000          & 0.528          & 1.000          & 0.996          & 0.989          & 0.998          & 0.000 & 0.000 & 0.106                         \\ \hline
\textbf{\#44247} & 0.188          & 0.062          & 0.426          & 0.995          & 0.988          & 0.998          & 0.028 & 0.077                         & 0.057                         \\ \hline
\textbf{\#1c96c} & 0.600          & 0.315          & 0.834          & 0.998          & 0.992          & 1.000          & 0.000 & 0.000 & 0.376                         \\ \hline
\textbf{\#e5ee5} & 0.667          & 0.358          & 0.883          & 0.998          & 0.992          & 1.000          & 0.000 & 0.000 & 0.377                         \\ \hline
\textbf{\#cd16c} & 0.857          & 0.496          & 0.983          & 0.998          & 0.992          & 1.000          & 0.000 & 0.000 & 0.378                         \\ \hline
\textbf{\#b3ca6} & 0.875          & 0.538          & 0.986          & 0.999          & 0.994          & 1.000          & 0.000 & 0.000 & 0.000 \\ \hline
\end{tabular}}
\caption{\label{tab:PNO}Performance of the proposed DLAD and assessed radiologists for the finding pneumothorax (PNO).}
\end{table}

Pneumothorax (PNO) is an infrequent diagnosis, with only 8 images (prevalence: 0.8\%) demonstrating abnormality. DLAD identified 7 of these as PNO+ (\textit{Se} of 0.875 (0.538-0.986)) and classified an additional 73 PNO- images as PNO+ (\textit{Sp} of 0.922 (0.903-0.938)). Considering the rarity of this condition, these outcomes are highly favorable, with only one radiologist (the most experienced) achieving similar results. The lower \textit{PPV} can be attributed to the higher rate of false positives and the low prevalence of PNO.

\newpage
\subsection{Forest Plots}

\begin{figure}[H]
\centering
\includegraphics[width=1\textwidth]{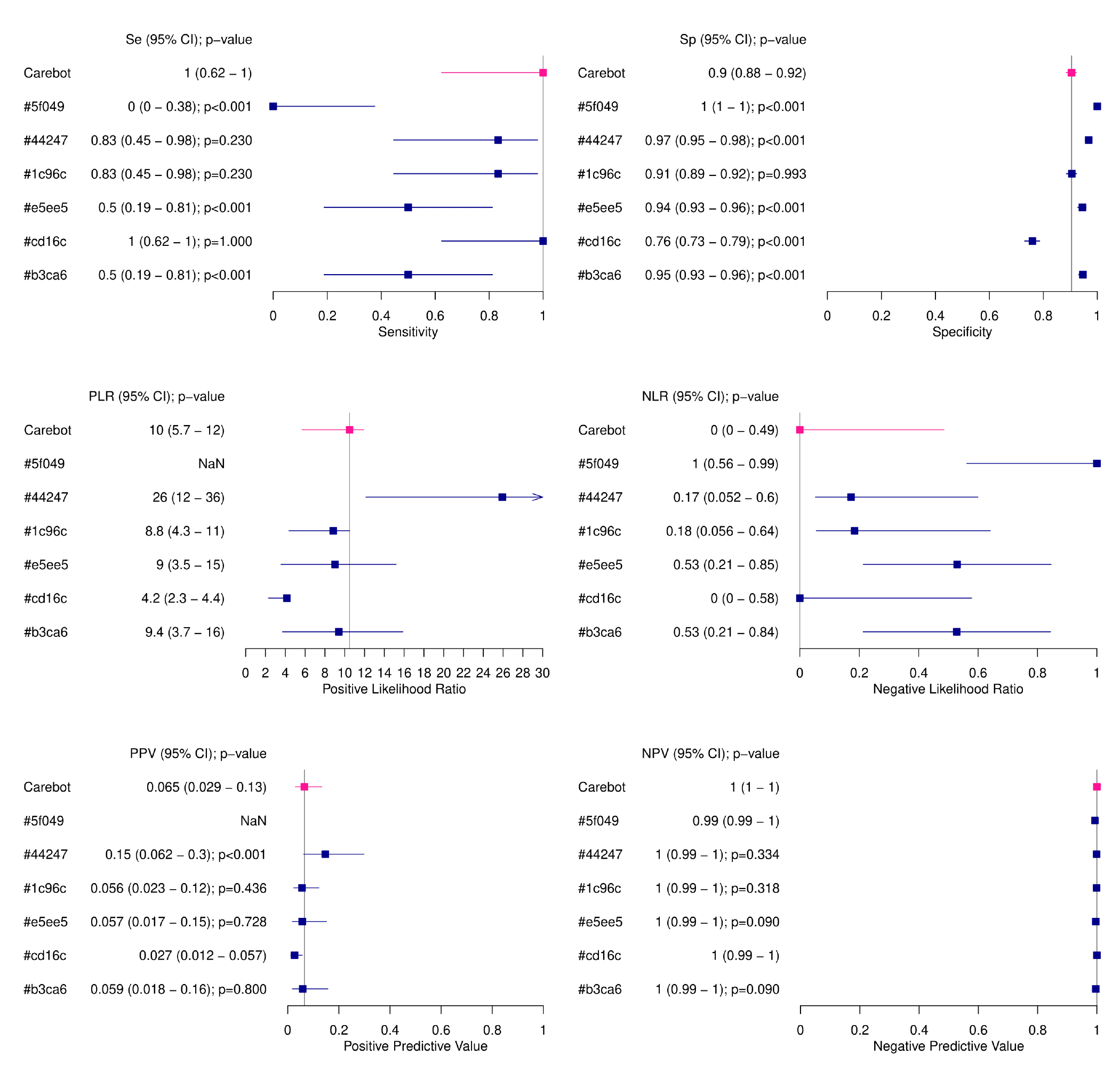}
\caption{\label{fig:ATE}\textbf\scriptsize{Forest plots for proposed DLAD and assessed radiologists for the finding atelectasis (ATE).}}
\end{figure}

\newpage

\begin{figure}[H]
\centering
\includegraphics[width=1\textwidth]{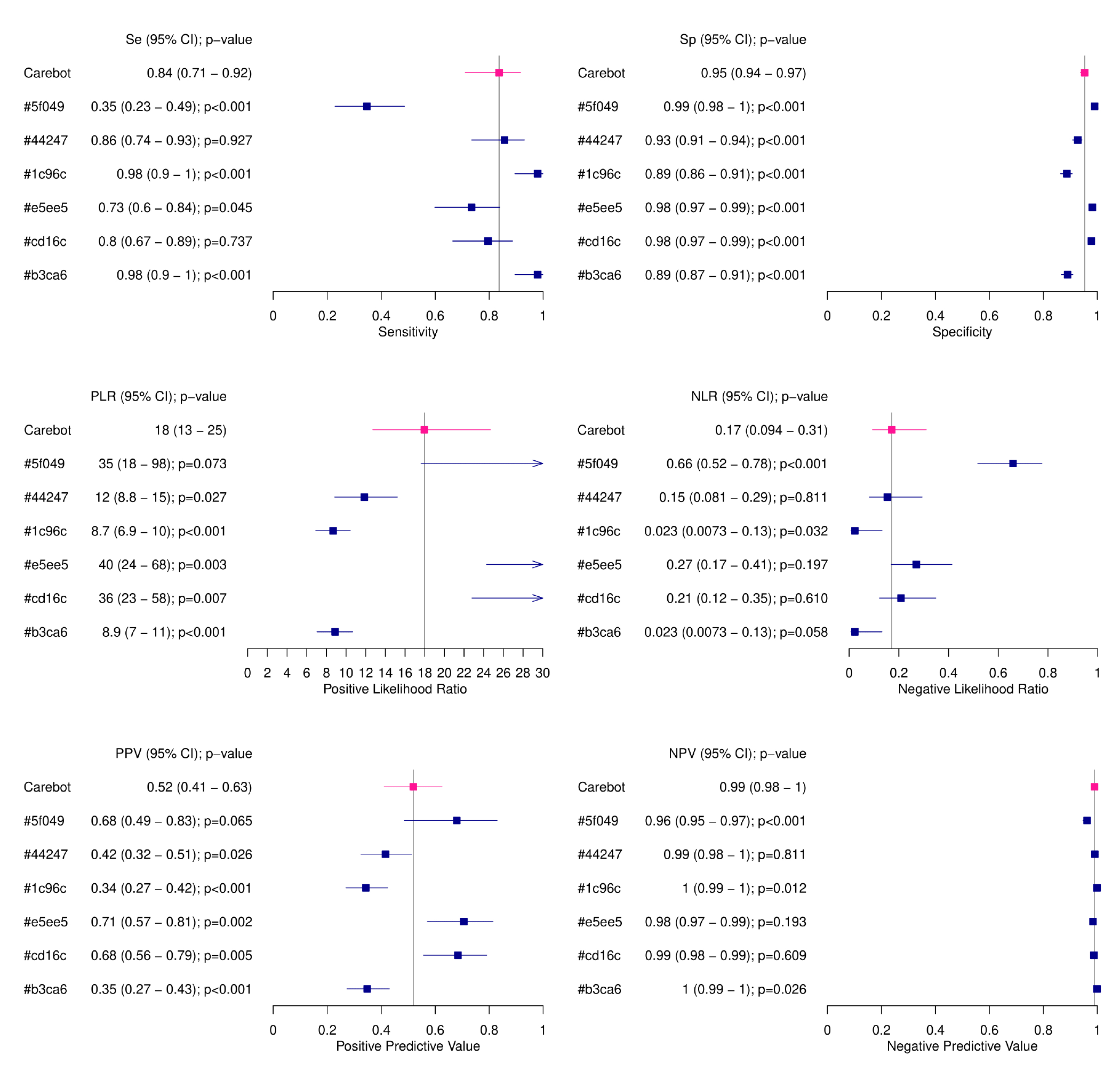}
\caption{\label{fig:CMG}\textbf\scriptsize{Forest plots for proposed DLAD and assessed radiologists for the finding cardiomegaly (CMG).}}
\end{figure}

\newpage

\begin{figure}[H]
\centering
\includegraphics[width=1\textwidth]{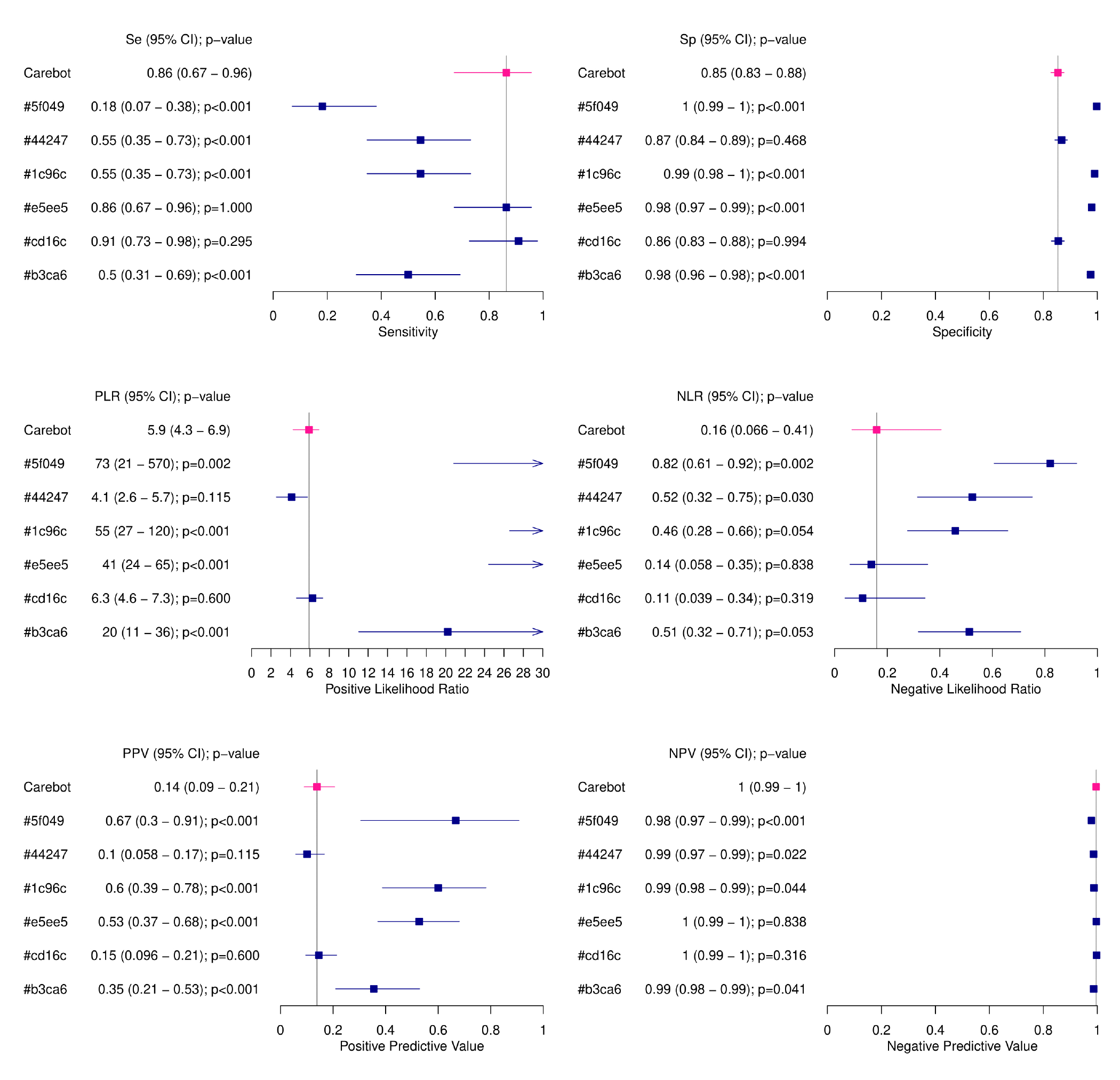}
\caption{\label{fig:CON}\textbf\scriptsize{Forest plots for proposed DLAD and assessed radiologists for the finding consolidation (CON).}}
\end{figure}

\newpage

\begin{figure}[H]
\centering
\includegraphics[width=1\textwidth]{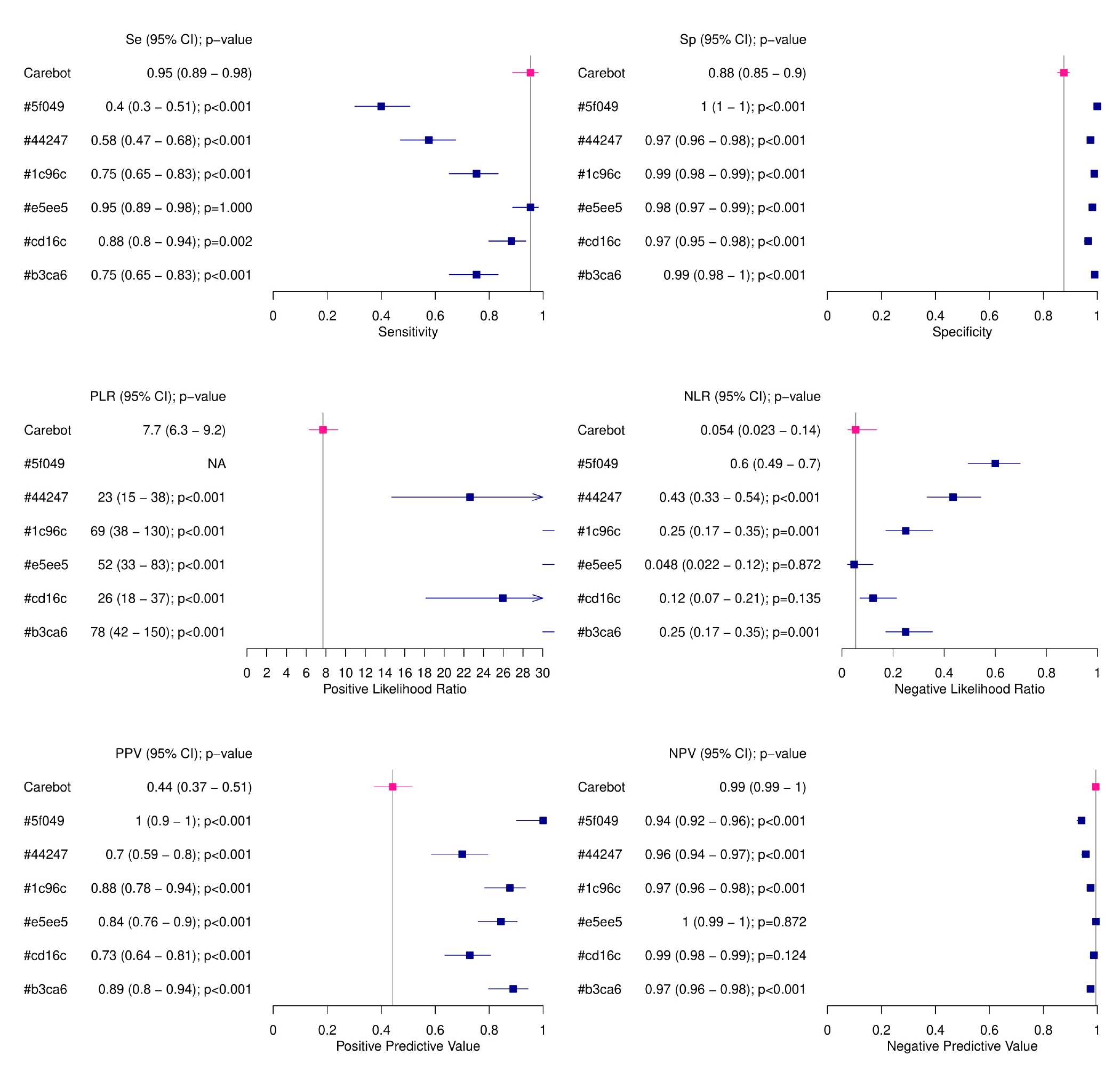}
\caption{\label{fig:EFF}\textbf\scriptsize{Forest plots for proposed DLAD and assessed radiologists for the finding pleural effusion (EFF).}}
\end{figure}

\newpage

\begin{figure}[H]
\centering
\includegraphics[width=1\textwidth]{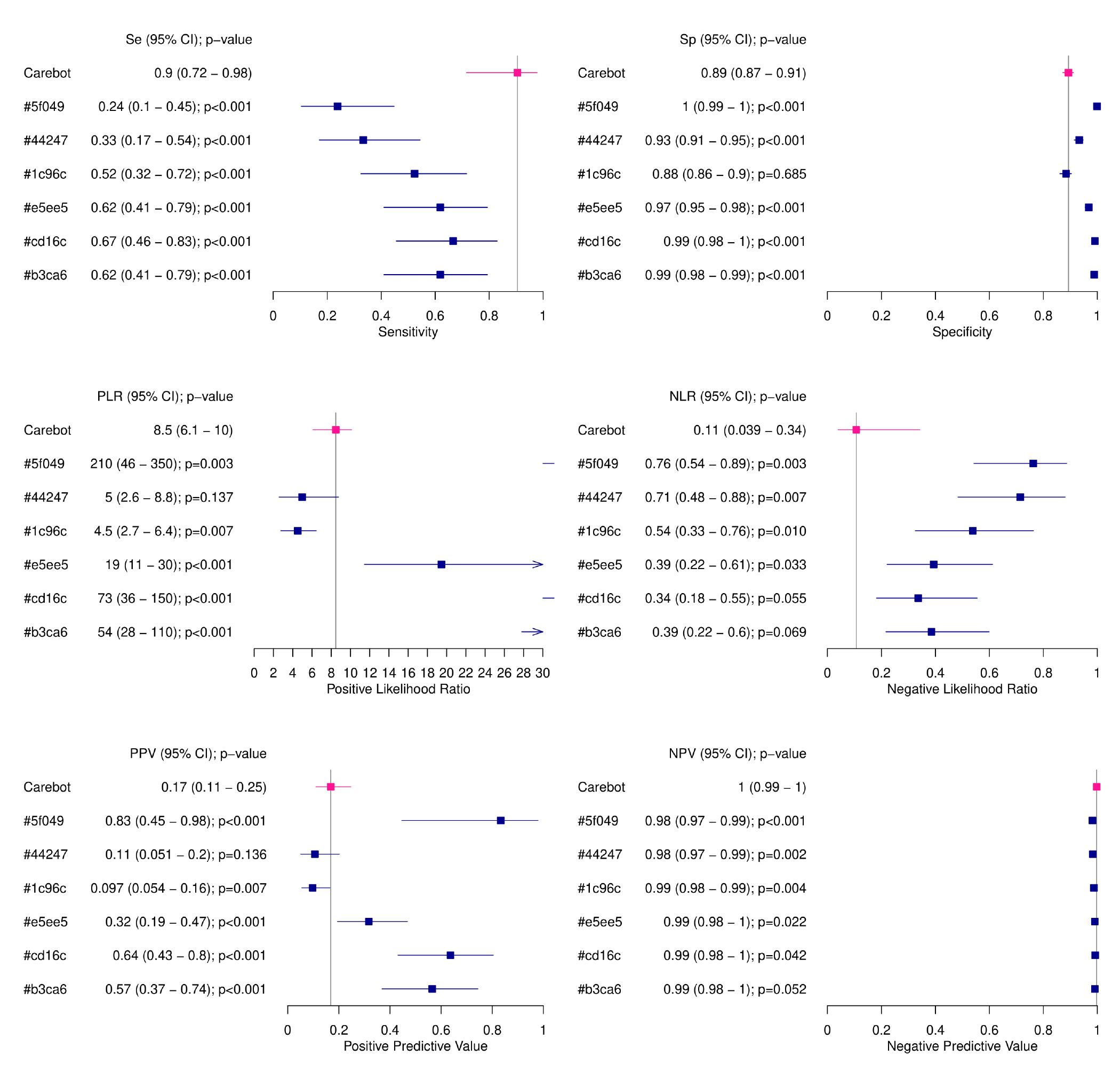}
\caption{\label{fig:LES}\textbf\scriptsize{Forest plots for proposed DLAD and assessed radiologists for the finding pulmonary lesion (LES).}}
\end{figure}

\newpage

\begin{figure}[H]
\centering
\includegraphics[width=1\textwidth]{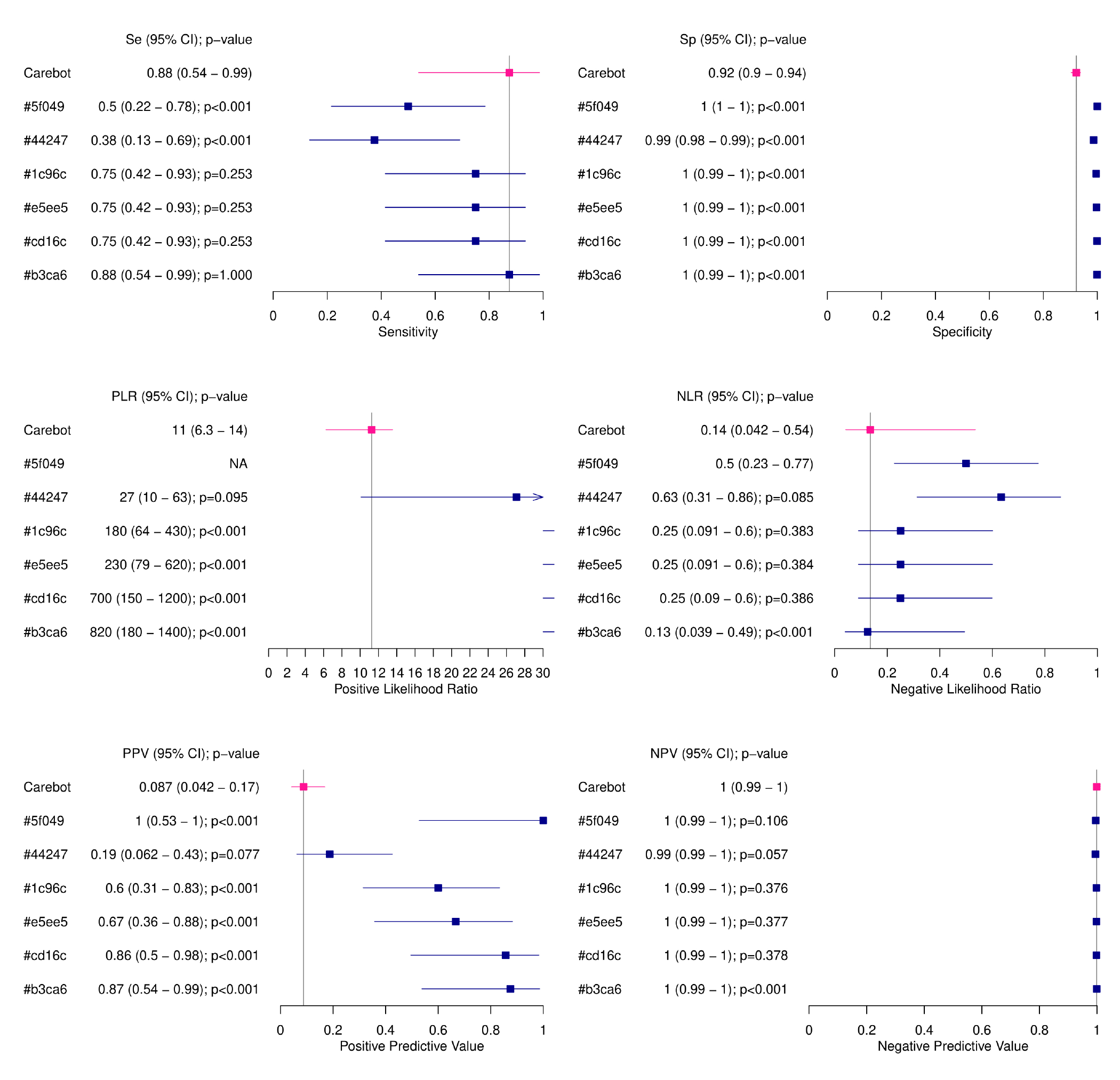}
\caption{\label{fig:PNO}\textbf\scriptsize{Forest plots for proposed DLAD and assessed radiologists for the finding pneumothorax (PNO).}}
\end{figure}

\newpage

\begin{figure}[H]
\centering
\includegraphics[width=1\textwidth]{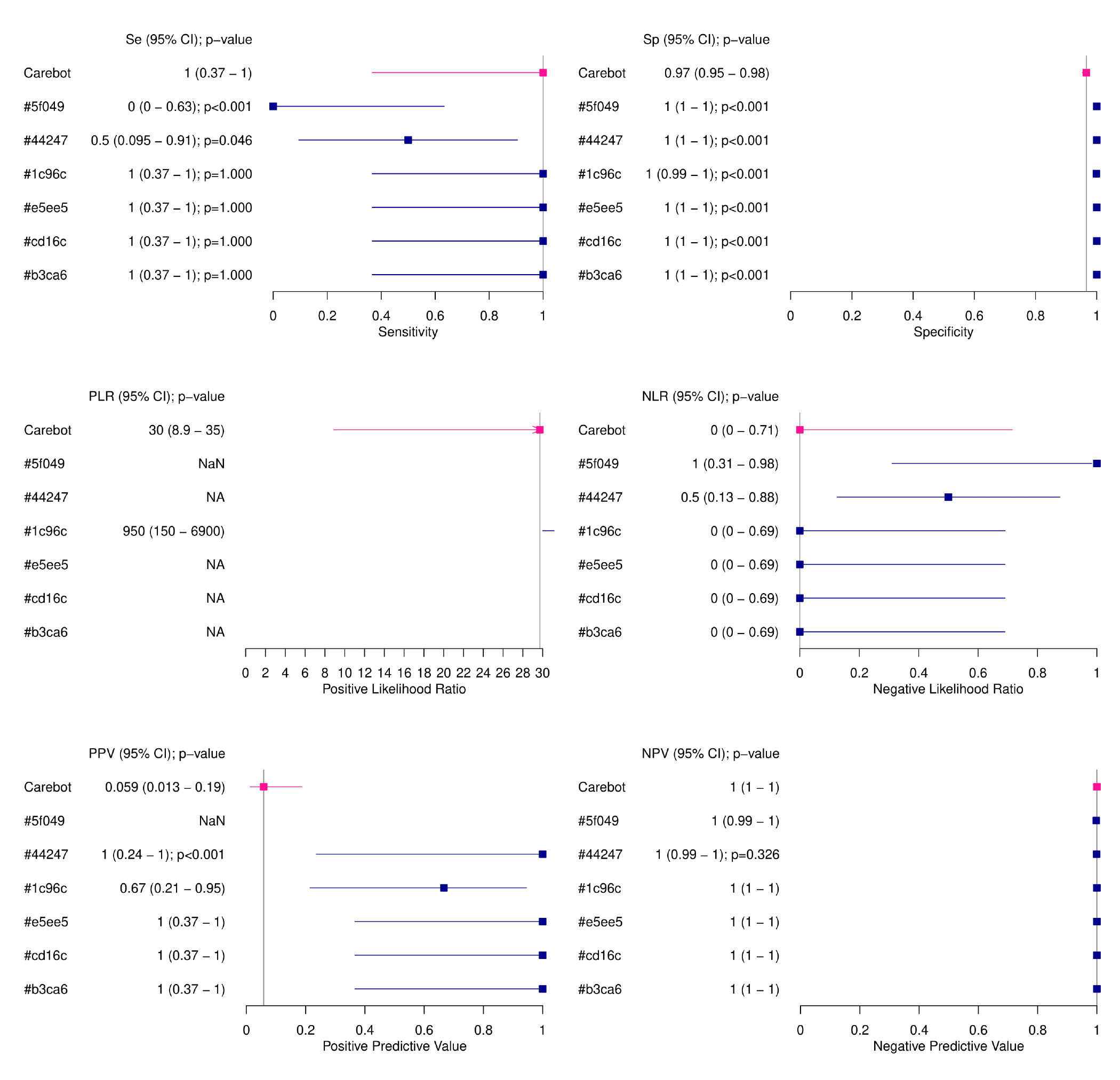}
\caption{\label{fig:SCE}\textbf\scriptsize{Forest plots for proposed DLAD and assessed radiologists for the finding subcutaneous emphysema (SCE).}}
\end{figure}

\newpage

\section{Conclusion}
The proposed DLAD (Carebot AI CXR) showed better or comparable results compared to the doctor's assessment, depending on the finding and its rarity as well as the experience of the doctor. It generally achieved higher sensitivity (\textit{Se}) and significantly higher specificity (\textit{Sp}) than expected, confirming the robustness of the model. The positive predictive value (\textit{PPV}) was generally rather lower (worse) than the doctor's assessment, due to the higher number of false positives, as a result of the setting of the algorithm itself to make DLAD classify questionable and suspicious images as abnormal. 

A large variability in the accuracy of the assessment of the findings was observed among physicians. In general, success rates correlated with physician experience. As a result, the proposed DLAD can be considered beneficial for both less and more experienced doctors. Only the very rare subcutaneous emphysema (SCE) was rated more favorably by radiologists, but given the very low prevalence of the finding, these are imprecise estimates. Pneumothorax (PNO) also showed very wide confidence intervals for the estimates, given the low prevalence. DLAD showed the most accurate classification for the finding of cardiomegaly (CMG) and pulmonary lesions (LES).

\subsection*{Abbreviations}
AI = Artificial Intelligence \\
ML = Machine Learning\\
DL = Deep Learning\\
CAD = Computer-Aided Diagnosis System\\
DLAD = Deep-Learning-based Automatic Detection Algorithm\\
Se = Sensitivity\\
Sp = Specificity\\
PPV = Positive Predictive Value\\
NPV = Negative Predictive Value\\
PLR = Positive Likelihood Ratio\\
NLR = Negative Likelihood Ratio\\
CI = Confidence Interval

\subsection*{Informed Consent Statement}
Informed consent was obtained from all subjects involved in the study.

\subsection*{Conflict of interest}
In relation to this study, we declare the following conflicts of interest: the research was funded by Carebot s.r.o.

\newpage

%
%

\bibliographystyle{plain}
\bibliography{main}

\end{document}